# Squash 2: A Hierarchicial Scalable Quantum Mapper Considering Ancilla Sharing

Mohammad Javad Dousti, Alireza Shafaei, and Massoud Pedram

Department of Electrical Engineering,

University of Southern California, 3740 McClintock Ave.

Los Angeles, CA 90089, USA

### Abstract

We present a multi-core *reconfigurable quantum processor* architecture, called *Requp,* which supports a hierarchical approach to mapping a quantum algorithm while sharing physical and logical ancilla qubits. Each core is capable of performing any quantum instruction. Moreover, we introduce a scalable quantum mapper, called *Squash 2*, which divides a given quantum circuit into a number of quantum modules—each module is divided into $k$ parts such that each part will run on one of $k$ available cores. Experimental results demonstrate that Squash 2 can handle large-scale quantum algorithms while providing an effective mechanism for sharing ancilla qubits.

## 1 Introduction

Mapping quantum circuits directly to a quantum fabric is a challenging task due to the gigantic size of quantum circuits. These circuits comprise of two parts: a netlist of *quantum logical operations* followed by the *quantum error correction* (QEC) circuit. The QEC increases the circuit size by one or two orders of magnitude depending on the decoherence degree and the desired fidelity of results. To handle this growth in the size, circuits are mapped in two levels. The lower-level mapping, which is done by the *physical-level mapper*, maps a universal set of quantum operations in a fault-tolerant fashion followed by an appropriate QEC circuit to a given *physical machine description* (PMD). In the higher-level mapping, which is performed by the *logical-level mapper*, the *logical* circuit is mapped to an abstraction of the PMD assuming that the universal set of fault-tolerant quantum operations is provided by the lower level. This approach addresses the increase in size by the QEC in the first level very well, but it does not help for the second level. Real-size quantum circuits (even without QEC) are so large that traditional mappers introduced by previous researchers cannot efficiently handle them [1].

Reference [2] shows that Shor's factorization algorithm for a 1024-bit integer has $1.35 \times 10^{15}$ physical instructions. Assuming that the one-level ⟦7,1,3⟧ Steane code is used in this implementation, each logical operation results in about $10^5$ physical instructions. Hence, this algorithm has almost $1.35 \times 10^{10}$ logical operations. As can be seen, the physical-level mapper can handle the low-level QEC in a reasonable time as the number of physical instructions is not so high ($\sim 10^5$ physical instructions) [3]. On the other hand, mapping $1.35 \times 10^{10}$ logical operations is very time consuming.

Fortunately, quantum circuits can be partitioned into repetitively-used quantum modules. This means that mapping one instance of these modules is sufficient. For instance, Fig. 1 shows the phase estimation algorithm which is the core of several well-known and useful quantum algorithms such as Shor's factorization algorithm [4] and quantum random walk [5]. As can be seen, in this circuit the controlled unitary is a module which is repeated $n$ times throughout the circuit with different exponents (in modules 2 to $n+1$). The exponent denotes the number of repetitions for the corresponding circuit. Clearly, identifying the quantum modules and avoiding the remapping can exponentially improve the mapping speed for this circuit.

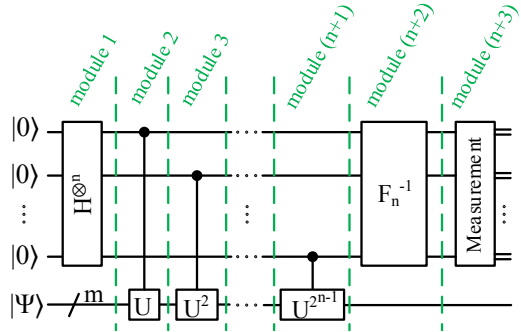

**Fig. 1.** Quantum circuit representation of the phase estimation algorithm [4]. Modules are identified in this circuit.

Another major stumbling block for realizing a scalable quantum computer is the limited amount of physical qubits. Each logical operation is implemented in a fault-tolerant manner based on the adopted QEC code, and using a certain amount of *physical data qubits* and *physical ancilla qubits*. Physical data qubits uniquely belong to their corresponding logical qubits, and hence cannot be shared. However, physical ancilla qubits, which are used to store intermediate information, may participate in the QEC circuit of various logical operations at different time instances. Similarly, *logical ancilla qubits* are used as scratch pads in the logical level, and can be reused in different quantum modules. This reuse of ancilla qubits is referred to as *ancilla sharing*. Escalating the ancilla sharing increases the latency of the entire circuit while saving the precious quantum resources and vice versa. This trade-off is similar to the well-known area-delay trade-off in VLSI circuits.

This paper introduces a novel quantum architecture, called *reconfigurable quantum processor architecture* (Requp), in order to address the problem of ancilla sharing. Requp has $k$ quantum cores each of which contains a *quantum reconfigurable compute region* (QRCR) and dedicated level one and two quantum caches and is surrounded by a quantum memory. Quantum cores are arranged on a 2-D mesh topology. Each QRCR has a constrained amount of physical ancilla qubits while trying to share this limited resource among several quantum operations so as to minimize the latency. The major contribution of this architecture lies in its reconfigurability where it supports quantum operations with different number of physical ancilla qubits. Moreover, it can be reconfigured to host various number of cores with different amounts of logical ancilla which enables logical ancilla sharing. These differences are quite substantial and neglecting them leads to over provisioning of quantum physical qubits.

Using the module extraction method and the proposed architecture (Requp) mentioned above, a scalable quantum mapper, called *scalable quantum mapper considering ancilla sharing* (*Squash 2*), is introduced. Squash 2 initially divides the given circuit into a number of quantum modules. For each module, it builds a *quantum module dependency graph* (QMDG) based on the data dependency among the operations and modules. QMDG is then partitioned into $k$ sub-graphs and bound to the quantum cores. These sub-graphs are subsequently scheduled and mapped to the Requp with $k$ quantum cores. Finally, result of mapping for each quantum module is combined in order to generate the entire mapping of the given circuit. The source code of Squash 2 can be obtained from http://sportlab.usc.edu/downloads.

Squash 2 which is covered in this paper extends the earlier version of Squash [6] in the following aspects:

- Considering Requp reconfiguration overhead to increase the accuracy for latency calculation.

- Adding another level of memory hierarchy in Requp to consider logical ancilla sharing. The earlier version only considered physical ancilla sharing.

- Utilizing hierarchical fault-tolerant quantum assembly (HF-QASM) language which permits to hierarchically map a quantum algorithm. The earlier version could only support one level of hierarchy in the quantum algorithm specification.

Moreover, this paper introduces a quantum mapping flow and demonstrates how Squash 2 fits into the flow. Besides, experimental results are extended by adding three new benchmarks with various sizes. Last but not least, the effect of hierarchical design on the Squash 2 performance is studied.

The remainder of this paper is organized as follows. Sections 2 explains the basics of quantum computing and the related work. Section 3 introduces a novel quantum mapping flow. Section 4 presents the new architecture (Requp), whereas Section 5 explains the proposed mapper (i.e., Squash 2). Experimental results are presented in Section 6, and finally Section 7 concludes the paper.

## 2 Quantum Computing Basics

### 2.1 Basics

A quantum bit, *qubit*, is a physical object (e.g., an ion or a photon) that carries data in quantum circuits. Qubits interact with each other through *quantum gates*. Depending on the underlying quantum computing technology, a universal set of quantum gates is available at the physical level. More precisely, each quantum fabric is natively capable of performing a universal set of one and two-qubit instructions (also called *physical instructions*). However, the importance of fault-tolerant quantum computation dictates the quantum circuits to be generated from *fault-tolerant* (FT) quantum operations. A universal (but redundant) set of FT operations includes $H$, $S$, $S^\dagger$, $T$, $T^\dagger$, $X$, $Y$, $Z$, and $CNOT$ operations [4], which may differ from physical instructions supported at the physical level. Fortunately, each FT quantum operation (or quantum operation for short) can be realized by using a composition of these physical instructions. Accordingly, a logical level circuit contains quantum operations where QEC is also applied.

A quantum circuit fabric is arranged as a 2-D array of identical *cells*. Each cell contains sites for creating qubits, reading them out, performing instructions on one or two physical qubits, and resources for routing qubits (or equivalently swapping their information to the neighboring qubit). In practice, however, an abstract *quantum architecture* (QA) is built which hides the physical information and the QEC details. Operation sites in this QA are capable of performing any quantum operation. The QA is also equipped with syndrome extraction circuitries following the quantum operation in order to prevent error propagation that may have been introduced by the quantum operation.

A *quantum compilation/synthesis* tool generates a reversible quantum circuit composed of quantum operations. Every qubit in the output circuit is called a *logical* qubit, which is subsequently encoded into several *physical* qubits in order to detect and correct potential errors on qubits. Physical qubits are comprised of two types: 1) *physical data qubits* and 2) *physical ancilla qubits*. Physical data qubits carry the encoded data of the logical qubits. Based on the type and the concatenation level of the QEC, a logical qubit is encoded to seven or more physical data qubits. On the other hand, physical ancilla qubits are used as scratchpads and can be *shared* among different logical qubits for the error correction procedure. Similarly, logical qubits have two kinds: 1) *logical data qubits* and 2) *logical ancilla qubits*. Logical data qubits carry crucial information throughout the runtime of program, whereas logical ancilla qubits can be reused.

A *high-level mapping* tool schedules, places, and routes the logical circuit on the QA. To achieve this, the quantum algorithm is initially modeled as a *quantum module dependency graph* (QMDG), in which nodes represent quantum operations or modules and edges capture data dependencies [1]. More precisely, module (or operation) $\theta_j$ depends on module (or operation) $\theta_i$ if $\theta_i$ and $\theta_j$ share at least one qubit and $\theta_j$ is the first module (or operation) after $\theta_i$ in the circuit that uses this (these) shared qubit(s). This dependency is shown as $\theta_i \rightarrow \theta_j$. For instance, Fig. 2 depicts an FT implementation of a three-input Toffoli operation [7] along with its QMDG.

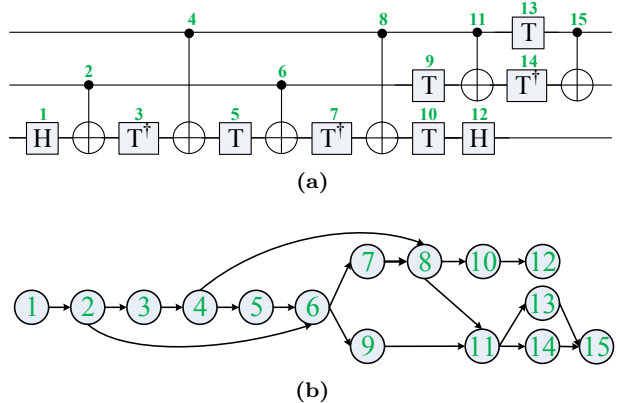

**Fig. 2.** (a) A fault-tolerant implementation of a three-input Toffoli gate [7]. (b) The corresponding QMDG where each node represents a circuit operation.

Next, the QMDG is mapped to the desired QA. The latency of the quantum algorithm mapped to the QA can be calculated as the length of the longest path (critical path) in the mapped QMDG, where the length of a path in the QMDG is in turn the summation of latencies of operations located on that path plus routing latencies of their qubit operands [1]. Critical path of the mapped QMDG may not be the same as the original QMDG, since the latter does not contain routing latencies. This can change the scheduling slacks, and hence may increase the critical path of the entire graph.

### 2.2 Prior Work

**Quantum Architectures:** Metodi *et al.* propose the first QA called *quantum logic array* (QLA) which is a 2-D array of super-cells called *tiles* [8]. Each tile comprises of an $n \times n$ array of cells so as a logical operation can fit in. Thaker *et al.* observe that the parallelism in quantum circuits is very limited [9]. Hence, they suggest *compressed QLA* (CQLA) which separates the array into two regions: *memory* and *compute*. In the memory region, the qubits which do not participate in any operation at the current time are stored. These qubits absorb less noise and hence require a lighter error correction scheme. In other words, the error correction needs fewer physical ancilla qubits for every physical data qubit (a ratio of 1 to 8). On the other hand, the qubits in the compute region actively participate in the quantum operations. Hence they require a much larger number of ancilla qubits. Since the compute region occupies much smaller area than the memory region, this new architecture helps save a lot of unnecessary physical ancilla qubits which are used in QLA. Memory region is also further broken down into the cache and the global memory to address the qubit locality issue required by the compute region.

**Quantum Mapping Techniques:** The quantum mapping problem, similar to the corresponding problem in the traditional VLSI area, is known as a hard problem. Whitney *et al.* suggest a CAD flow for mapping a quantum circuit fault-tolerantly to an ion-trap fabric [2]. To address the scalability issue, they adopt the two-level (physical and logical) mapping. Other levels of hierarchy are handled manually without any automation. Jones *et al.* propose a five-layer stack for implementing a quantum computer [10]. This work does not show how to overcome the complexity of the "*logical layer*" and tries to address other complexities in the design by adding more layers. In [1], we have suggested to use a quick estimation method called *LEQA* to calculate the circuit latency instead of a full-fledged mapping. Even though this approach is quite fast, it does not provide the detailed mapping. Moreover, it requires a flattened high-level netlist as the input which requires a huge amount of disk space to store the netlist and a large memory in order to store its data structures. Additionally, LEQA does not consider the ancilla sharing problem. Although several heuristics have been proposed in the literature for solving the quantum mapping problem, none of them is able to deal with large circuits [2, 8, 9, 11, 12].

## 3 Design Flow

The entire tool chain flow of a quantum mapper is depicted in Fig. 3. This flow is depicted in the *Enhanced Functional Flow Block Diagram* (EFFBD) representation, which is detailed in [13]. The purpose of the EFFBD is to indicate the sequential relationship of all functions that must be accomplished by a system. EFFBDs show the same tasks identified through functional decomposition and display them in their logical and sequential relationship. We use the following symbols and notation:

- Functions are represented as rectangular boxes (blocks) with an associated label number, and data flow overlay for capturing data dependencies is represented by arrows.

- Each function receives a unique label number that can be used as identification, and is shown in the top-left corner of the function.

- The primary inputs to the flow are depicted in green, the intermediate results are shown in blue, and primary outputs are in orange.

These tasks consider QEC to protect qubits against noise, and *quantum control* (QC) to increase the fidelity. As a result, a combination of quantum algorithm, PMD, QEC code, and QC protocol acts as the main inputs to a design flow. More accurately, the primary inputs to the design flow are classified as follows:

- **Scaffold code:** A quantum algorithm written in the Scaffold language, which is a quantum programming language described in [14].

- **Tech. parameters:** A description of the PMD, the QC protocol, and the QEC code.

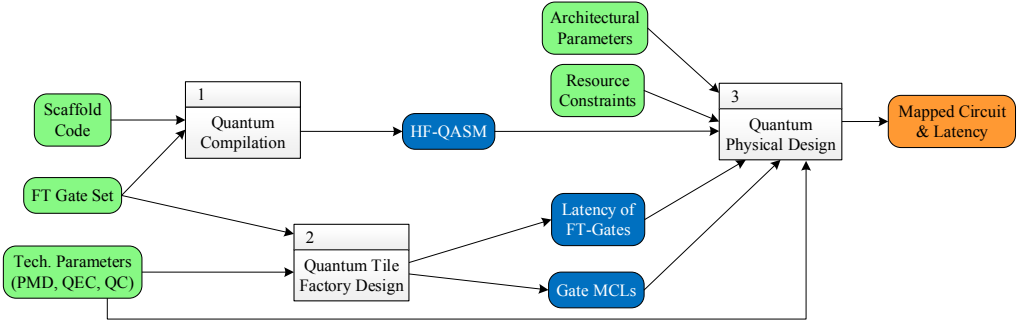

**Fig. 3.** Quantum mapping tool chain flow.

- **FT gate set:** A universal set of quantum gates to be used in a fault-tolerant quantum computation. This universal set is referred to as *quantum gate set* (QGS). For the given QEC code, the exact implementation of each of the gates in the QGS is given. In addition to the standard gates, new gates can also be added to this set as long as an efficient, fault-tolerant implementation for the new gates exists in the given QEC code, and correct costs are considered in the synthesis step.

- **Architectural parameters:** Specification of the quantum architecture.

- **Resource constraints:** Physical ancilla qubits are limited resources that should be determined for the tool chain.

The description of functional blocks in Fig. 3 are as follows.

1. "1- Quantum Compilation" receives the Scaffold code and the QGS, compiles the code, and synthesizes it to a circuit consisting of gates from the QGS. It is assumed that either the given Scaffold code is modular or the compiler decomposes it to various modules (which is outside the scope of this paper). The output of this block is called HF-QASM, which will be formally defined later in this section.

2. The generated HF-QASM is fed to "3- Quantum Physical Design" in order to be mapped to the quantum circuit fabric. In addition, this block calculates the circuit latency (i.e., the number of physical time steps to execute the quantum algorithm) which is the main quality measure for the design flow.

3. Prior to the physical design, a *machine control language* (MCL) code for each gate in the QGS is generated in the "2- Quantum Tile Factory Design", which uses (i) the realization of each gate in QGS using native instructions of the given PMD by considering the QC protocol, and (ii) encoding/decoding circuits for the given QEC code in order to calculate the latency of each FT gate. In addition, the latency of performing each FT gate is derived.

Based on the above discussion, it can be seen that the proposed flow uses a *meet-in-the-middle approach* to generate the MCL code. On one hand, the top-down development in the front-end tool suite breaks the Scaffold code into gates that are chosen from the QGS. On the other hand, the bottom-up development in the tile factory designer tool generates optimized MCL codes (in terms of gate or qubit counts) for each gate of the QGS. These two paths meet each other in the back-end, where high-level gates are scheduled, placed and routed on the 2D tiled architecture, and then each high-level gate as well as move operation are replaced by appropriate MCL codes in order to build the circuit MCL. Note that the path that goes through the quantum tile factory design tool is an offline process that executes independently of the quantum algorithm.

A key benefit of the meet-in-the-middle approach is that the efficiency of the top-down (forward) synthesis approach and the accuracy of the bottom-up physical design approach are simultaneously realized. Additionally, the design tool scalability is ensured by hiding the physical implementation details of logic operations (and moves) in a given PMD for given QC and QEC protocols from the algorithm and code developers. Ease of verification of the design by separating front-end transformations from gate library verification is another benefit.

In order to address the scalability issue, quantum compiler should generate its output in hierarchical form. In other words, the final output should be modular such that repetitive code appears once. This not only manages the output size, but also helps the physical design block to map each part of the code once. Thus, we developed the *hierarchical fault-tolerant quantum assembly* (HF-QASM) language.

Fig. 4 shows the HF-QASM grammar. Every program written in HF-QASM is required to have a *main module*. This module calls other modules in the program. Other modules in turn can call the others. The only limitation is that no circular module call is allowed. The qubits defined at the top of the main module are logical data qubits. These qubits are passed to other modules through parameters. On the other hand, qubits that are defined at the top of non-main modules are logical ancilla qubits. Their life-time is limited to the execution of their parent module as opposed to logical data qubits which exist for the duration of the whole program execution. This grammatical distinction enables the quantum mapper to effectively share logical ancilla qubits among the modules. Fig. 5 demonstrates the implementation of a Fredkin gate in FT-QASM using three Toffoli gates. As can be seen, Toffoli gates are implemented based on the FT gate decomposition presented in Fig. 2(a).

```
name → [a-z,A-Z]+[a-z,A-Z,0-9]*
num → [1-9][1-9]*
whitespace → \n | \r | \t |
comment → # .* (\n | <EOF>)
```

(a)

```
start → (module)* main
main → module main { body }
module → module name ( param_list ) { body }
param_list → param (,param)*
param → qubit (*)? name
body → (def;)+ (gate;)+
def → qubit ([num])? name
gate → (one_qubit_gate | two_qubit_gate | call)
one_qubit_gate → (H|X|Y|Z|S|S†|T|T†|Prep0|MeasX|MeasY|MeasZ) (arg)
two_qubit_gate → CNOT (arg, arg)
call → name (call_list)
call_list → arg (, arg)*
arg → name | name[num]
```

(b)

**Fig. 4.** (a) Definitions of HF-QASM tokens. (b) HF-QASM grammar. Regular expression is used to simplify the grammar specification. Note that the star character (**\***) used for the *param* variable is a *terminal*. It determines whether a parameter is an array or not (similar to the C language).

The focus of this paper is on the "3- Quantum Physical Design" block of Fig. 3. In order to prepare intermediate inputs to our tool, we use *Scaffold Compiler* [15] as the quantum compilation tool to generate HF-QASM and our prior work QUFD [3] to provide

```
module Toffoli(qbit c1, qbit c2, qbit t){
    H (t);
    CNOT (c2, t);
    Tdag (t);
    CNOT (c1, t);
    T (t);
    CNOT (c2, t);
    Tdag (t);
    CNOT (c1, t);
    T (c2);
    T (t);
    CNOT (c1, c2);
    H (t);
    T (c1);
    Tdag (c2);
    CNOT (c1, c2);
}
module main(){
    qbit a[3];
    #a[0] is control and a[1] and a[2] are target qubits
    Toffoli(a[0], a[2], a[1]);
    Toffoli(a[0], a[1], a[2]);
    Toffoli(a[0], a[2], a[1]);
}
```

**Fig. 5.** Description of a Fredkin gate in HF-QASM.

the latency and MCL of FT gates for the ion-trap technology. Note that Squash is not limited to a particular quantum technology; however, the ion-trap technology is selected since it is among the most promising methods for realizing quantum circuits to date [16].

## 4  Proposed Architecture

The CQLA architecture reviewed earlier assumes that the number of required physical ancilla qubits for all of the logical operations followed by the QEC is the same. Hence, CQLA accounts for a certain amount of physical ancilla qubits for every logical operation in the compute region. However, an important subset of logical operations, called *non-transversal* operations, requires more ancilla than *transversal* operations. It has been proven that every universal logical operation set contains at least one non-transversal gate which varies based on the employed QEC [17]. Table 1 summarizes the physical ancilla requirements for two typical QEC codes and various logical operations. As can be seen, a non-transversal operation requires half an order of magnitude (in the *Steane code*) up to more than one order of magnitude (in the *Bacon-Shor code*) more ancilla qubits compared to that of transversal operations. Moreover, a two-qubit transversal operation (like CNOT) requires twice ancilla qubits compared to that of a one-qubit transversal operation.

With this observation, the compute region cannot be a pre-allocated area with a fixed number of physical ancilla qubits for all of operations; otherwise, it leads to an overestimation of the required ancilla. Hence, we propose the *quantum reconfigurable compute region* (QRCR) which distributes the ancilla qubits in the compute region based on dispatched operations. In other words, physical ancilla qubits are shared among the operations which are being executed based on their ancilla qubit requirements. To further speed up the computation and eliminate the overhead of qubit routing, a hierarchical memory design is adopted. The first level of the hierarchy is the *quantum L1 cache* which stores qubits that are immediately needed after the execution of current operations in the QRCR. The second level is the *quantum L2 cache* which keeps qubits required in the rest of the current quantum module computation. Using this hierarchy, the routing delay overhead can be mostly hidden. More precisely, the routing delay is substantially smaller than the delay of logical operations, because the routing involves qubit movement (or information swap) which can be done directly by using fast primitive operation(s), whereas logical operations require time consuming QECs. Considerable routing delays are the time required to load the qubits from the quantum cache to the QRCR and the time to transform Requp architecture so as to execute a new quantum module (more on this later).

**Table 1.** Ancilla requirements for various QEC codes and operations.

| QEC | Operation Type | Operation | # of Ancilla Qubits |
|---|---|---|---|
| ⟦7,1,3⟧ Steane Code | Transversal | X, Y, Z, H, S, S† | 28 |
| | | CNOT | 56 |
| | Non-Transversal | T, T† | 100 |
| ⟦9,1,3⟧ Bacon-Shor Code | Transversal | X, Y, Z, H | 18 |
| | | CNOT | 36 |
| | Non-Transversal | S, S† | 58 |
| | | T, T† | 309 |

*Quantum memory* is the last level of memory hierarchy which holds qubits that are not necessary in the current quantum module but required by other quantum modules. This includes both logical data and ancilla qubits. Ancilla qubits can be stored in memory temporarily and used when they are needed. This enables Requp to provide logical ancilla sharing.

Fig. 6(a) depicts a *quantum core* which is comprised of a QRCR and a two-level quantum cache. As can be seen, QRCR is located at the center and surrounded by the quantum L1 cache followed by the quantum L2 cache. The highly shaded areas inside the QRCR have higher number of ancilla, whereas lightly shaded areas contain fewer ancilla. The arrangement of ancilla changes during the runtime of a quantum algorithm based on the operations being executed.

In large-scale algorithms, the size of the cache and the memory may grow. This increases the qubit routing delay which was already hidden by the long delay of logical operations. To avoid this effect, we further extend the quantum core architecture to the <u>*reconfigurable quantum processor architecture*</u> (Requp). A Requp contains multiple reconfigurable quantum cores which are connected to each other by *quantum interconnects*. Quantum interconnects are physically implemented similar to the rest of the quantum physical fabric. Here, this distinction is made for clarity. A quad-core Requp is shown in Fig. 6(b).

## 5 Squash 2

This section introduces a hierarchical scalable quantum mapper, which considers ancilla sharing, called *Squash 2*. Squash 2 adopts Requp as its underlying fabric abstraction. According to the design flow explained in Section 3, Squash 2 inputs can be classified as follows:

- **HF-QASM:** A quantum algorithm description specified in the HF-QASM language

- **Latency of FT-Gates & gate MCLs:** MCL and latency of each operation in the QGS

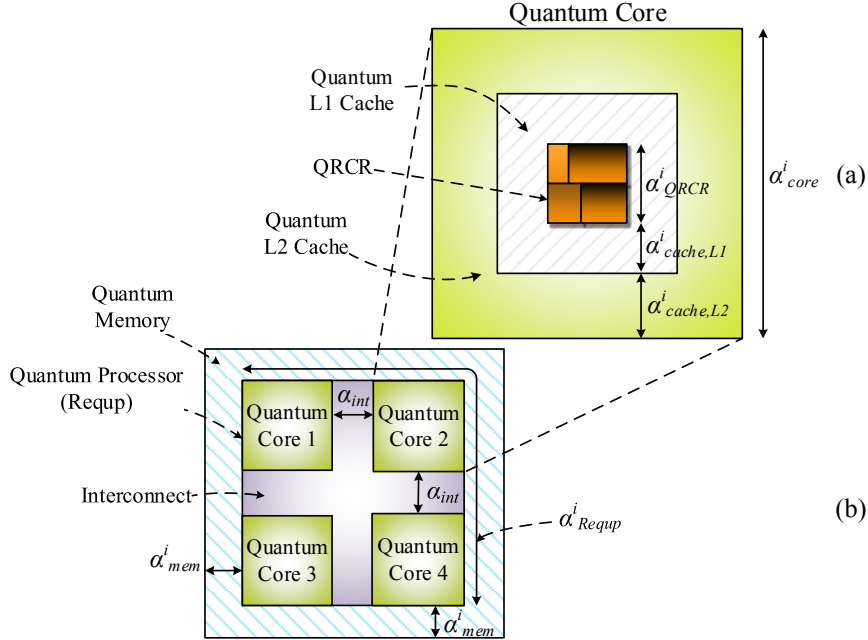

**Fig. 6.** (a) Structure of a quantum core. (b) Structure of a quad-core Requp surrounded with quantum memory. $\alpha_{int}$, $\alpha^i_{QRCR}$, $\alpha^i_{core}$, $\alpha^i_{cache,L1}$, $\alpha^i_{cache,L2}$, $\alpha^i_{mem}$, and $\alpha^i_{Requp}$ are the width of their respective components. Their detailed definitions are listed in Table 2.

- **Tech. parameters:** A QEC code description similar to Table 1 and a PMD-specific timing parameter which defines the delay of a qubit traveling the distance of a grid cell (called *qubit one-step delay* and denoted by $\beta_{PMD}$)

- **Architectural parameters:** The number of quantum cores ($k$), the interconnect width ($\alpha_{int}$), and a coefficient which models the effect of the L2 cache size on the routing speed ($\gamma_{cache,L2}$)

- **Resource constraints:** The total physical ancilla budget assigned to all of QRCRs ($B_{QRCR}$)

The output of Squash 2 is a circuit mapped to the designated fabric. The total circuit latency can be found from the final mapping solution.

As it is explained previously, early work found that quantum algorithms offer limited parallelism [9]. By investigating various quantum algorithms, including the phase estimation algorithm which is at the basis of several well-known and useful quantum algorithms (such as Shor's factorization algorithm [4] and quantum random walk [5]), we realized that quantum algorithms can be divided into major *modules* which cannot be run in parallel, i.e., they should be executed serially. The main reason is due to the *no-cloning theorem*, which does not allow a qubit to be replicated. This limitation forbids any fan-out in a quantum circuit. As a result, scheduling of modules becomes a trivial task— they should be run serially. Moreover, these modules in turn may also hierarchically contain a number of repetitively-used quantum modules. Mapping only one instance of these modules significantly reduces the mapping runtime overhead. On the other hand, because of the sequential execution of modules, logical ancilla may be shared among them.

Algorithm 1. presents the key steps involved in Squash 2. The first line of Algorithm 1. makes a QMDG for the given HF-QASM. Next, the QMDG is traversed in post-order

(i.e., first children, and then their parent), and nodes are stored in an ordered list called $\mathcal{S}$. After that, in the for-loop block (lines 4 to 9), the algorithm maps each of the modules separately. The final solution ($\mathcal{M}_m$) is the mapping of the root node in the QMDG (i.e., the *main* module) which is the $m^{th}$ element in the post-order traversal. This contains information about the mapping of logical operations to quantum cores as well as logical qubit movement. However, an MCL code deals with physical-level instructions and qubits. Hence, logical operations (including move or swap operations) in $\mathcal{M}_m$ are replaced by corresponding MCLs in order to produce the final physical-level mapping solution, denoted by $\mathcal{M}_m^{final}$ (line 11).

The for-loop body works as follows. Line 4 generates a QMDG called QMDG$_i$ as explained in Section 2. Next, QMDG$_i$ is broken into $k$ parts such that $k$ quantum cores can execute these parts simultaneously while having the minimum amount of inter-core communication (line 5). The routing delay matrix is then calculated, which comprises of the qubit routing delays between every pair of quantum cores (line 6). Each part is then bound (line 7) and mapped (line 8) to a quantum core, and finally combined (line 9). In the remainder of this section, the details of these steps (i.e., the for-loop body) are explained.

---

**Algorithm 1.** Squash 2

**Input:** An HF-QASM, QGS MCLs and their latencies, a QEC code, architectural parameters (i.e., number of quantum cores ($k$), qubit one-step delay ($\beta_{PMD}$), interconnect width ($\alpha_{int}$), and L2 cache size effect on the routing coefficient ($\gamma_{cache,L2}$)), and resource constraint (i.e., total physical ancilla budget assigned to all of QRCRs ($B_{QRCR}$))

**Output:** Mapped circuit of the given HF-QASM

1: Make a QMDG for the given HF-QASM
2: Traverse QMDG in the post-order way and store the nodes in an ordered list called $\mathcal{S} = <\mathcal{S}_1, \ldots, \mathcal{S}_m>$
3: **for** i=1 **to** m **do**
4:     Generate a QMDG for the operations & modules in $\mathcal{S}_i$ (called QMDG$_i$)
5:     $K$-way partition QMDG$_i$ to get $\mathcal{P}_i = \{\mathcal{P}_{i,1}, \ldots, \mathcal{P}_{i,k}\}$
6:     Calculate the routing delay matrix **d**
7:     Bind each $\mathcal{P}_{i,j}$ to one of the quantum cores
8:     Map each $\mathcal{P}_{i,j}$ to the designated quantum core
9:     Combine the mappings for all of the cores to derive $\mathcal{M}_i$
10: **end for**
11: Use FT-gate MCLs and $\mathcal{M}_m$ to generate final mapping called $\mathcal{M}_m^{final}$
12: **return** $\mathcal{M}_m^{final}$

---

### 5.1 QMDG $K$-Way Partitioning

A standard $k$-way partitioning algorithm takes a graph, and divides its node set into $k$ disjoint parts such that the parts are balanced in terms of their size and a minimum number of edges are cut. Using this method, the same workload is assigned to each quantum core, while inter-core communication is minimized. However, there is no guarantee that parts can be executed in parallel which is in fact a desired metric in order to reduce the runtime. As an example, consider the QMDG shown in Fig. 2(b), and assume a two-way partitioning is needed. A standard graph partitioning algorithm may suggest the dashed cut in Fig. 7(a) which partitions the graph into two parts with almost equal number of nodes. Unfortunately, this solution does not allow any parallelism. On the other hand, consider the dotted cut in Fig. 7(b). Even though one part has twice as many nodes as the other one in this partitioning, it is a better solution as parts can be executed simultaneously.

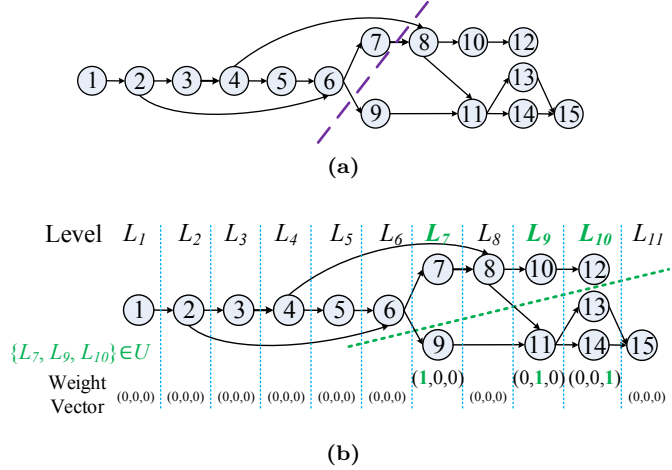

**Fig. 7.** (a) The standard two-way partitioning on the QMDG of a three-input Toffoli gate. (b) The MCGP partitioning with detailed steps for the same QMDG.

In order to guide the partitioning algorithm to produce parts that can be run in parallel, we employ the technique proposed in [18] which adopts the *multi-constraint graph partitioning* (MCGP) method explained in [19]. The MCGP method assigns an ($n_{con}$)-dimensional weight vector to each node, and then balances the total sum of the weight values among the parts in each dimension while minimizing the edge cut.

The weight vector for each QMDG node is calculated as follows. Initially, the QMDG is levelized. Let $n_i$ be the number of nodes at level $i$ (denoted by $L_i$), $U = \{L_i \mid n_i \geq k\}$, and $n_{con} = |U|$ (i.e., the number of levels that contain greater than or equal to $k$ nodes.) Then, weight vectors of size $n_{con}$ are assigned to each node. For nodes that are at level $L_i \notin U$, the weight vector is set to the zero vector. For other nodes, we first assign a label to each level $L_i \in U$ using the one-hot coding scheme. This label will be used as the weight vector for all of the nodes within the same level. Hence, by using one-hot coding, a unique dimension of the weight vector is assigned to all nodes at level $L_i \in U$. Therefore, the MCGP method is forced to partition these nodes into distinct parts so that the total weight in the corresponding dimension for each part is balanced.

**Example:** In order to perform the two-way MCGP on the QMDG shown in Fig. 2(b), the QMDG should be first levelized. The result of levelization determines $U = \{L_7, L_9, L_{10}\}$. Consequently, $n_{con} = 3$. All of the nodes except the ones associated with levels $L_7, L_9, L_{10}$ are assigned a zero vector (i.e., $(0,0,0)$). Using the one-hot coding, nodes 7 and 9 are assigned vector $(1,0,0)$, nodes 10 and 11 are labeled with vector $(0,1,0)$, and finally nodes 12, 13, and 14 are marked with vector $(0,0,1)$. This process is depicted in Fig. 7(b). An MCGP solver tries to separate nodes in the levels listed in $U$ to achieve a weight-balanced result. The dashed line shown in Fig. 7(b) is a desired partitioning solution with total weight of $(1,1,1)$ and $(1,1,2)$ for the top and bottom partitions, respectively. On the other hand, the solution given in Fig. 7(a) results in $(1,0,0)$ and $(1,2,3)$ weight vectors for two parts which is clearly not-balanced.

## 5.2 Requp Characterization

In this phase, based on the information obtained from the partitioning step, the quantum core is characterized in order to find accurate qubit routing delays between each pair of cores. Note that it is not necessary to use the same quantum core configuration for all of the quantum modules, because it is just an abstraction to simplify the mapping and to hide the technology details.

For this purpose, as shown in Fig. 6, five parameters, namely $\alpha^i_{QRCR}$, $\alpha^i_{core}$, $\alpha^i_{cache,L1}$, $\alpha^i_{cache,L2}$, and $\alpha^i_{mem}$ for the $i^{th}$ module are initially calculated. The approach is to derive the number of physical qubits that each area should be able to accommodate and then the desired parameters are calculated accordingly. In other words, we assume that Fig. 6 is a grid of cells, where each cell can accommodate only a qubit. We summarized the notations used in this subsection in Table 2.

**Table 2.** Requp nomenclature.

| Symbol | Definition |
|---|---|
| $\alpha_{int}$ | Interconnect width which is an input parameter (see Fig. 6) |
| $\alpha^i_{QRCR}$ | QRCR width for module $i$ (see Fig. 6) |
| $\alpha^i_{core}$ | Quantum core width for module $i$ (see Fig. 6) |
| $\alpha^i_{cache,L1}$ | Quantum L1 cache width for module $i$ (see Fig. 6) |
| $\alpha^i_{cache,L2}$ | Quantum L2 cache width for module $i$ (see Fig. 6) |
| $\alpha^i_{mem}$ | Quantum memory width for module $i$ (see Fig. 6) |
| $\alpha^i_{Requp}$ | Half perimeter length of the quantum processor for module $i$ |
| $B_P$ | Total physical ancilla budget |
| $B_{QRCR}$ | Total physical ancilla budget assigned to all of QRCRs |
| $D^{total}_L$ | Total logical data required throughout the quantum program |
| $D^i_L$ | Total logical data required by module $i$ |
| $L^i_{max}$ | Maximum number of logical qubits (data+ancilla) required by any core in module $i$ |
| $A^i_L$ | Number of logical ancilla that module $i$ requires |
| $A^i_{min}$ | Minimum physical ancilla requirement among quantum operations |
| $A^i_{max}$ | Maximum physical ancilla requirement among quantum operations |
| $Q^i_{mem}$ | Number of physical qubits reside in the memory of module $i$ |
| $d_{x,y}$ | Routing delay between core $x$ and core $y$ |
| $n_{x,y}$ | Manhattan distance between the center of core $x$ and $y$ |
| $\beta_{PMD}$ | Qubit one-step delay (technology dependent) |
| $\gamma_{cache,L2}$ | L2 cache size effect on the routing coefficient |
| $l_{code}$ | QEC code length |

$\alpha^i_{QRCR}$ can be obtained by

$$\alpha^i_{QRCR} = \left\lceil \sqrt{\frac{B_{QRCR}/k}{A^i_{min}}.l_{code} + B_{QRCR}/k} \right\rceil. \tag{1}$$

The first summation term in this equation accounts for the maximum number of physical data qubits that QRCR may host, whereas the second term accounts for the physical ancilla qubits. Multiplication by $l_{code}$ converts the number of logical qubits to the number of physical qubits. Note that $B_{QRCR}/k$ is the physical ancilla budget per QRCR. $B_{QRCR}$ should be chosen such that

$$k.A^i_{max} \leq B_{QRCR}. \tag{2}$$

In other words, there should be enough physical qubits in each core such that any operation which requires the maximum number of physical ancilla qubits (i.e., $A^i_{max}$) can also be executed. As an example, for the Steane code listed in Table 1, $A^i_{max}$, $A^i_{min}$, and $l_{code}$ are equal to 100, 28, and 7, respectively.

$\alpha^i_{core}$ is determined by

$$\alpha^i_{core} = \left\lceil \sqrt{\lceil \tfrac{9}{8} L^i_{max} \rceil \cdot l_{code} + B_{QRCR}/k} \right\rceil. \tag{3}$$

Note that a core should be able to host logical qubits (first term) as well as physical ancilla qubits (second term). $L^i_{max}$ can be calculated by referring to the partitioned set of operations for each core. Furthermore, $L^i_{max}/8$ logical ancilla qubits are reserved for the error correction of logical qubits in two levels of cache. As mentioned earlier, for the QEC of every eight logical data qubits in the cache or the memory, only one logical ancilla qubit is enough (see [9] for details.)

As suggested in [9], $\alpha^i_{cache,L1}$ can be set such that the L1 cache area becomes twice as large as the QRCR area. Hence, $\alpha^i_{cache,L1}$ can be calculated as

$$\alpha^i_{cache,L1} = min\left\{\left\lceil \frac{\sqrt{3}-1}{2} \alpha^i_{QRCR} \right\rceil, \frac{\alpha^i_{core} - \alpha^i_{QRCR}}{2}\right\}. \tag{4}$$

A minimum value is calculated in order to avoid over provisioning of resources for the L1 cache, i.e., the L1 cache plus QRCR area should not be larger than the area of the core.

Next, $\alpha^i_{cache,L2}$ can be derived based on the values of $\alpha^i_{QRCR}$, $\alpha^i_{cache,L1}$, and $\alpha^i_{core}$:

$$\alpha^i_{cache,L2} = \left\lceil \frac{\alpha^i_{core} - \alpha^i_{QRCR}}{2} - \alpha^i_{cache,L1} \right\rceil. \tag{5}$$

Using these four parameters, the communication delay for routing a qubit from the QRCR of core $x$ to the QRCR of core $y$ can be calculated as

$$d_{x,y} = \begin{cases} n_{x,y}\left(\alpha^i_{core} + \alpha_{int}\right)\beta_{PMD}, & x \neq y \\ \frac{(\alpha^i_{QRCR} + \alpha^i_{cache,L1} + \gamma_{cache,L2} \cdot \alpha^i_{cache,L2})}{2}\beta_{PMD}, & x = y \end{cases}. \tag{6}$$

The first case ($x \neq y$) considers the inter-core routing delay, whereas the second case ($x = y$) accounts for the delay of transferring a qubit from the cache (L1 or L2) into the QRCR. Coefficient $\gamma_{cache,L2}$ ensures the proper contribution of the L2 cache size to the routing delay of a qubit. In other words, if the L2 cache size becomes large enough, then the routing delay cannot be overshadowed by the long operation delay, and hence should be considered in the routing delay calculation. We capture this effect with the $\gamma_{cache,L2}$ coefficient.

The quantum memory includes logical and physical ancilla qubits which are not necessary for the mapping of module $i$ and are reserved for other modules. Their count can be calculated as

$$Q^i_{mem} = \left((\max_i\{A^i_L\} - A^i_L) + \lceil \tfrac{9}{8}(D^{total}_L - D^i_L) \rceil\right) \cdot l_{code}. \tag{7}$$

The term $\max_i\{A^i_L\}$ is the maximum number of logical ancilla any module may use in the current quantum program. Thus, the first term considers logical ancilla qubits that are not currently being used, stored in the memory, and will be shared with other modules. Note that unlike logical data quibts, logical ancilla qubits do not require error corrections (because they are used as scratch pads). The second term captures logical data qubits along with their error correction qubits. Accordingly, $\alpha_{mem}^i$ is determined as

$$\alpha_{mem}^i = \left\lceil \sqrt{(\alpha_{Requp}^i)^2 + 2Q_{mem}} - \alpha_{Requp}^i \right\rceil, \tag{8}$$

where $\alpha_{Requp}^i$ is the half perimeter length of the quantum processor and can be defined as

$$\alpha_{Requp}^i = (n_{1,k} + 1) \cdot \alpha_{core}^i + n_{1,k} \cdot \alpha_{int}. \tag{9}$$

Total physical ancilla qubits that are used to run a qunatum program can be calculated as follows.

$$B_P = B_{QRCR} + \max_i \left\{ (\lceil \tfrac{1}{8}(A_L^i + D_L^i) \rceil + \lceil \tfrac{1}{8}(D_L^{total} - D_L^i) \rceil) \cdot l_{code} \right\} \tag{10}$$

Note that the first term captures the budget assigned to QRCRs for performing quantum operations, whereas the max term accounts for the maximum ancilla required for error correction of logical data qubits in memory or caches among all modules. The first term inside the max operator considers QEC qubits used inside two-level caches and the second term represents QEC qubits required for logical data qubits which reside in the memory. As can be seen, the value of $B_P$ depends on the quantum program and hence we consider $B_{QRCR}$ as the input to Squash 2.

## 5.3 Resource Binding

After partitioning the QMDG, the resultant parts should be bound to the quantum cores such that the total routing delay of qubits between cores is minimized. Since the scheduling of the QMDG is not known at this step, we cannot focus on minimizing the total routing delay of the operations on the critical path. Furthermore, the scheduling requires this binding information in order to properly schedule two dependent operations assigned to two different quantum cores.

The binding problem can be formulated as follows.

$$Min \sum_{m=1}^{k} \sum_{n=1}^{k} \sum_{x=1}^{k} \sum_{y=1}^{k} a_{m,n} \cdot a_{x,y} \cdot d_{n,y} \cdot w_{m,x} \tag{11}$$

subject to

$$\sum_{n=1}^{k} a_{m,n} = 1, \qquad \text{for } 1 \leq m \leq k, \tag{12}$$

$$\sum_{m=1}^{k} a_{m,n} = 1, \qquad \text{for } 1 \leq n \leq k, \tag{13}$$

where $a_{m,n}$ is a binary variable, which is 1 if $\mathcal{P}_{i,m}$ is bound to quantum core $n$ and 0 otherwise, and $w_{m,x}$ denotes the number of qubits that traverse from part $\mathcal{P}_{i,m}$ to $\mathcal{P}_{i,x}$. The objective function (11) is the sum of inter-core communication delays while constraints (12) and (13) ensure a one-to-one assignment between parts and quantum cores. Since $k$ is small, the computation time to solve the resulting *0-1 quadratic program* (0-1 QP) is of little concern.

## 5.4 Requp reconfiguration overhead

During the execution of a quantum module, another module may be called. By traversing the QMDG derived from an HF-QASM in post-order way, we ensure that Squash 2 knows how to map callee modules. However, the architecture requires to be reconfigured to be prepared for a callee module. This imposes some delays and has to be taken into account. Similarly, this reconfiguration is required when mapping of a module is finished, and another module is about to be mapped.

The reconfiguration delay is of the qubit routing delay type and it depends on the distance qubits are required to travel. The reconfiguration delay has two components:

1. **Architecture transformation delay:** This delay is due to the variation of Requp parameters for different cores explained in previous subsections. The qubits are required to be routed in correct locations to form the new Requp configuration.

2. **Logical ancilla qubits routing delay:** Similar to the previous case, the new module may require additional logical ancilla which has to be routed from memory to desired locations.

These two delay components occur in parallel, thus the total reconfiguration delay between operations/modules $x$ and $y$ is the maximum of them and is shown by $T_{x,y}^{rec}$ in this paper.

## 5.5 Mapping

The objective of scheduling the QMDG on $k$ quantum cores is to minimize the overall latency while ensuring that the number of physical ancilla qubits used in each quantum core is no more than the given budget ($B_{QRCR}$). The aforesaid scheduling problem is similar to the well-known *minimum-latency resource-constraint multi-cycle* (MLRC-MC) scheduling problem [20] in high-level synthesis with the following difference. The MLRC-MC problem does not deal with the cost of moving data among resources whereas in our formulation the resources (i.e., quantum cores) lie on a given grid, and therefore, their average communication costs can be pre-calculated (see Eq. (6) and the previous subsection). More precisely, our problem formulation is as follows.

$$Min \; \mathcal{L} \quad (14)$$

subject to

$$\sum_{\theta_x \in \mathcal{P}_{i,j}} \sum_{y=0}^{T_x-1} u_{x,z-y} A_{\theta_x} \leq B_{QRCR}/k, \quad 1 \leq z \leq \mathcal{L}_{init}, 1 \leq j \leq k, \theta_x \text{ is an operation} \quad (15)$$

$$\sum_{\theta_x \in \mathcal{P}_i} \sum_{y=0}^{T_x-1} u_{x,z-y} \leq 1, \quad 1 \leq z \leq \mathcal{L}_{init}, \theta_x \text{ is a module} \quad (16)$$

$$\sum_{y=1}^{\mathcal{L}_{init}} u_{x,y} = 1, \quad \forall \theta_x, \quad (17)$$

$$S_x + T_x + T'_{x,y} \leq S_y, \quad \theta_x \rightarrow \theta_y, \quad (18)$$

$$S_x + T_x - 1 \leq \mathcal{L}, \quad \forall \theta_x \text{ without any successors}, \quad (19)$$

where $\mathcal{L}$ is the total number of scheduling levels, $\theta_x$ represents an operation or a module in the QMDG, $u_{x,y}$ is a binary variable which is 1 if $\theta_x$ is scheduled to start at scheduling level $y$ and 0 otherwise, $A_{\theta_x}$ denotes the physical ancilla requirement of $\theta_x$, $\mathcal{L}_{init}$ is an upper bound for the total number of scheduling levels ($\mathcal{L}$), $S_y$ is equal to the scheduling level where $\theta_y$ is scheduled, i.e., $u_{y,S_y} = 1$, $T_x$ is the delay of $\theta_x$, and $T'_{x,y}$ is defined as

$$T'_{x,y} = \begin{cases} d_{m,n}, & \theta_x \in C_m \text{ and } \theta_y \in C_n \\ T^{rec}_{x,y}, & \text{otherwise} \end{cases}, \quad (20)$$

which means that $T'_{x,y}$ is equal to the routing delay between core $m$ and core $n$ if $\theta_x$ and $\theta_y$ are operations (not modules) and are bound to quantum cores $m$ and $n$, respectively. If either of $\theta_x$ or $\theta_y$ is a module, then an architecture reconfiguration is required; thus, the reconfiguration delay ($T^{rec}_{x,y}$) should be considered.

Eq. (15) sets a constraint on the total number of physical ancilla that each core can use at each scheduling level. On the other hand, Eq. (16) ensures that at most only one module is scheduled at each scheduling level. This constraint is required to enable hierarchical mapping and isolation among hierarchies. Eq. (17) ensures that all of the operations and modules are scheduled. Eq. (18) makes sure that dependent operations/modules are properly scheduled, i.e., an operation/module starts after its predecessor in the QMDG is finished. Eq. (19) assures that the operations/modules in the last scheduling level are scheduled to finish their execution before or at the scheduling level $\mathcal{L}$. We modified the list scheduling method presented in [21] as described above to solve the scheduling problem.

Using the Requp architecture, the logical and physical ancilla sharing problems are solved during the scheduling. Moreover, the placement problem has already been solved in the prior step (i.e., resource binding step). Additionally, as it is explained earlier, the routing delay is hidden by the operation delay. Hence, a simple routing algorithm like the xy-routing fits well for the purpose of transferring qubits (or equivalently swapping their information) through the interconnection network of Requp.

## 6 Experimental Results

Squash 2 is developed in Java. It uses METIS 5.1 [19] as the partitioning engine and Gurobi 6 [22] for solving the 0-1 QP. We set a 60 sec timeout for solving each 0-1 QP. If the optimum solution could not be found within this time period, Gurobi would pick the best solution that has been found by the end of the timeout period. This is especially useful for the cases where $k > 8$. A desktop computer with an Intel Core i7-3770 CPU running at 3.40 GHz and 8 GB of memory is employed for the experiments.

The $[\![7,1,3]\!]$ Steane code with the information presented in Table 1 is adopted as the QEC code. Moreover, Requp parameters are set as $\alpha_{int} = 3$ and $\gamma_{cache,L2} = 0.2$. Besides, $\beta_{PMD}$ is set to $10\mu s$.

Table 3 summarizes the benchmarks that we used to evaluate Squash 2. It also compares the size of the compiled codes in QASM and HF-QASM formats. As can be seen, for complex algorithms such as *Triangle Finding Problem*, QASM format is completely inefficient because it requires more than 60 GB of disk space. On the other hand, HF-QASM can be successfully adopted.

In the rest of this section, first the latency-ancilla count trade-off in quantum circuits is studied using Squash 2. Then, the optimum number of quantum cores for above-mentioned benchmarks is found. After that, the resource requirement of Requp, CQLA, and QLA are analytically compared. Finally, Squash 2 is compared with the state-of-the-art mapper.

### 6.1 Investigating the latency-ancilla count trade-off

As it is explained earlier, physical ancilla qubits are precious resources in quantum computers. Increasing the total ancilla budget lowers the circuit latency and vice versa. In order to study this effect using Squash 2, we start with two implementations of the *Binary*

**Table 3.** Summary of benchmarks used in this paper.

| Benchmark | Ref. | Module Count | Problem Size | File Size | | Size Ratio |
|---|---|---|---|---|---|---|
| | | | | QASM | HF-QASM | |
| Grover's Algorithm | [23] | 6 | n=100[†] | 548 KB | 52 KB | 11 |
| | | 6 | n=300[†] | 3.2 MB | 160 KB | 20 |
| | | 6 | n=500[†] | 7.0 MB | 268 KB | 27 |
| Binary Welded Tree (3M-BWT) | [24] | 3 | $n = 43, 3 \leq s \leq 19$[‡] | 5.52 MB – 34.9 MB | 373 KB – 377 KB | 15 – 95 |
| Binary Welded Tree (HM-BWT) | | 35 | | | 84 KB – 88 KB | 67 – 406 |
| Ground State Estimation | [25] | 170 | M=6[§] | 481 MB | 200 KB | 2,462 |
| Triangle Finding Problem | [26] | 332 | n=5[*] | 63 GB | 504 KB | 129,418 |
| | | 10,202 | n=10[*] | Failed[¶] | 28 MB | N/A[¶] |

[†] A database of $2^n$ elements is being searched.
[‡] $n$ is the height of the tree and $s$ is a time parameter within which the solution is found.
[§] $M$ is the molecular weight of a molecule.
[*] $n$ is the number of nodes in the graph.
[¶] The QASM file size exceeded 75GB and Scaffold compiler was crashed while generating the output file.

*Welded Tree* (BWT) algorithm as the benchmark. The first implementation has only three modules (shown by *3M-BWT*), whereas the other one (which is highly modular) has 35 modules (denoted by *HM-BWT*.) For every given QRCR physical ancilla budget ($B_{QRCR}$), the number of quantum cores ($k$) is set to the value which gives the lowest latency.

The trade-off between latency and the ancilla budget ($B_{QRCR}$) is shown in Fig. 8. As can be seen, Squash 2 produces better results for 3M-BWT compared to HM-BWT. The reason is that 3M-BWT has only three modules compared to HM-BWT which has 35 modules. This allows Squash 2 to utilize parallelism inside large modules of 3M-BWT and perform better partitioning. As it will be demonstrated later, high modularity of input files results in faster mapping runtime.

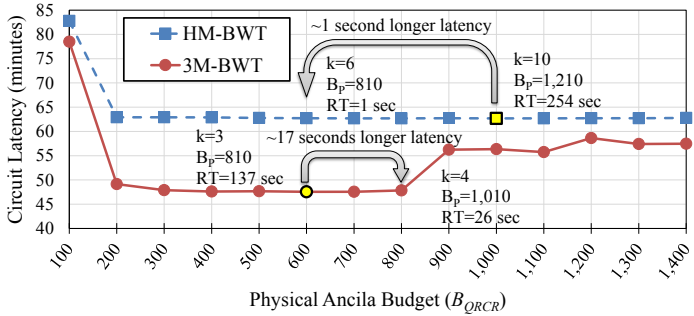

**Fig. 8.** Latency-ancilla count trade-off for the Binary Welded Tree (BWT) algorithm. RT represents the runtime of Squash 2.

Fig. 8 also shows that the latency of HM-BWT is saturated for $B_{QRCR} \geq 200$, whereas for 3M-BWM, the latency is lowered when $200 \leq B_{QRCR} \leq 800$ and then it increases when $800 < B_{QRCR}$. The low latency of 3M-BWT is due to the high parallelism of large modules. When the ancilla budget is increased, the size of cores grows which results in greater inter-core communication. Thus, the circuit latency increases.

Moreover, Fig. 8 depicts the optimum $B_{QRCR}$ value in yellow dots for two benchmarks. Gray arrows show how the runtime of Squash 2 can be improved with small sacrifice in the circuit latency (i.e., 1 second for HM-BWT, and 17 seconds for 3M-BWT). As can be seen for HM-BWT, the runtime reduction results in 400 fewer physical ancilla qubits, whereas for 3M-BWT, it comes at the cost of 200 more ancilla qubits. Generally, Squash 2 performs the fastest when $5 \leq k \leq 8$. For smaller values of $k$, the partitioning step takes longer, because the length of the weight vector for partitioning is larger. On the other hand, for larger values of $k$, the resource binding step is the runtime bottleneck.

## 6.2 Finding the optimum number of quantum cores

The optimum value for quantum core count ($k$) varies based on the parallelism inside a given circuit. Fig. 9 shows the latency of 3M-BWT and HM-BWT as a function of quantum core count. For every $k$, we tried various QRCR physical ancilla budget values and picked the one which results in the best (lowest) latency. As can be seen, for HM-BWT, for $k \geq 2$ the latency saturates and becomes minimum when $k = 10$. On the other hand, 3M-BWT has low latency when $1 \leq k \leq 4$ and it becomes minimum when $k = 3$. Note than in the case when $k = 1$, Squash 2 allows parallelism when $B_{QRCR}$ is sufficiently large. In other words, several operations can be executed in parallel inside a core. In contrast, in Fig. 8 when $B_{QRCR} = 100$, the architecture can only accomodate one core (i.e., $k = 1$) due to the constraint (2) which results in the highest latency. Again, gray arrows show the runtime reduction. Similar to Fig. 8, HM-BWT enjoys not only from runtime reduction but also lower ancilla usage. On the other hand, runtime reduction results in higher ancilla usage for 3M-BWT.

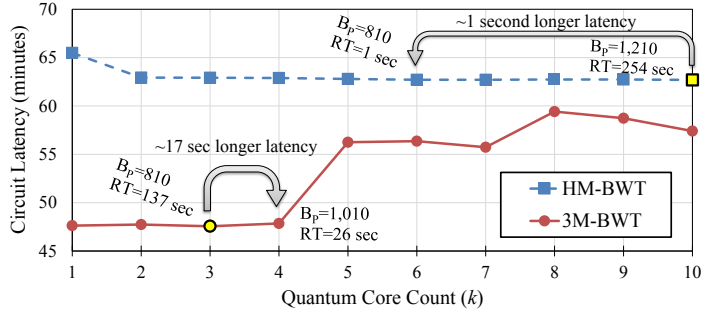

**Fig. 9.** Finding the optimum number of quantum cores for the Binary Welded Tree (BWT) algorithm. RT represents the runtime of Squash 2.

Fig. 10 shows the optimum $k$ value for various benchmarks. Note that the vertical axis is scaled differently for each chart and also the time unit varies. *Grover's Algorithm* with various sizes achieves its minimum latency when $k = 10$. The latency of *Ground State Estimation* algorithm reduces significantly when the number of cores goes beyond one. Last but not least, the algorithm for solving the *Triangle Finding Problem* is the largest of all benchmarks. It takes about a year for the small-size problem ($n = 5$) and about 170 years for the larger one ($n = 10$) to finish. Note that these algorithms might be of practical use when a dramatic speed-up for the underlying quantum technology is achieved. For instance, the delay of primitive operations for the ion-trap technology is about ten to hundreds of microseconds; however, the quantum dot technology lowers this delay by three orders of magnitude [16]. Unfortunately, this technique still suffers from high-error rate and is not as mature as the ion-trap technology.

Gray arrows in Fig. 10 show how runtime of Squash 2 can be reduced with minimum change in the latency. In all of benchmarks, the runtime reduction also reduces the physical ancilla requirement. Moreover, the latency increase is negligible in Grover's Algorithm and Ground State Estimation; however, for Triangle Finding Problem, the latency increases by 3 hours and 13 days when $n = 5$ and $n = 10$, respectively.

## 6.3 Resource usage comparison among Requp, CQLA, and QLA architectures

In the QLA architecture, every qubit requires to be placed in a quantum tile. Each tile needs to support all types of quantum operations and their respective QEC codes. Hence, in the case of one-level ⟦7,1,3⟧ Steane code, the required physical ancilla in this architecture is equal to 100×(total qubit count). CQLA limits this value to the maximum number of parallel operations the architecture should be able to execute. For instance, if $z$ parallel operations are supported (which is significantly smaller than the total qubit count), $100 \times z$ ancilla qubits are required. Requp improves this resource limitation by considering the fact that all of the parallel operations may not require the maximum number of ancilla qubits (i.e., 100). Therefore, Requp allows to run at most $(100/28) \times z$ operations at the same time while still having the same worst case parallelism as CQLA. This discussion reveals that Requp performs more efficiently in the average case compared to CQLA and behaves as bad as CQLA in the worst case.

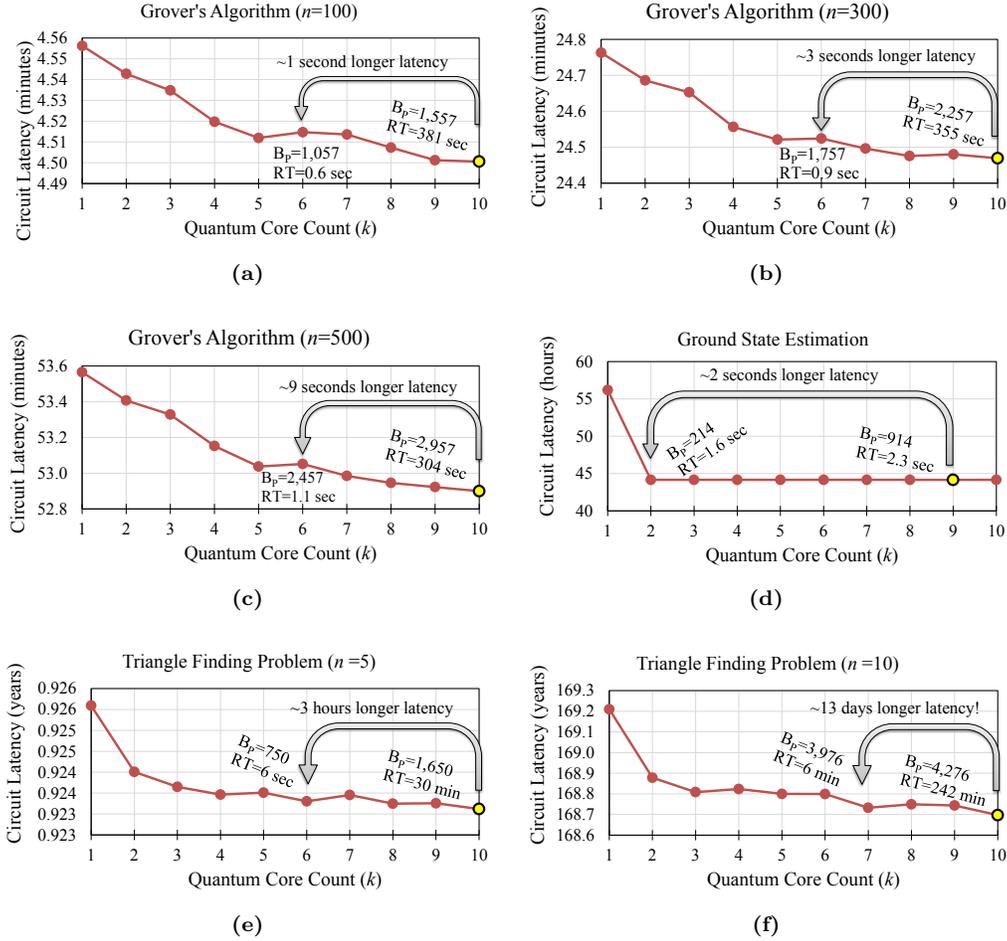

**Fig. 10.** Finding the optimum core count for various benchmarks given that the optimum physical ancilla budget has already been found. The optimum point is shown in yellow. Note that the vertical axis is scaled differently for each chart plus the time unit varies. RT represents the runtime of Squash 2.

### 6.4 Comparison between Squash 2 and QSPR

In this section, the performance and the quality of results produced by Squash 2 is compared with that of *QSPR* which is introduced in [12]. QSPR is a full-fledged quantum mapper which is recently improved to support the QLA architecture [1]. Unfortunately,

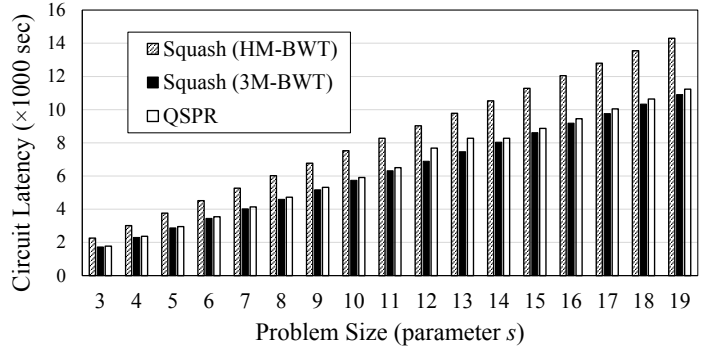

**Fig. 11.** Comparison of BWT latency mapped by QSPR and Squash 2.

no quantum mapper for the CQLA architecture is available to the public for comparison.

For comparing Squash 2 and QSPR, we compiled various sizes of the BWT algorithm based on a parameter called $s$. This parameter is varied from 3 to 19, where $s = 19$ is the problem size of interest. (For the previous experiments, $s$ was set to 5.) Again, here we consider both of HF-QASM implementations for BWT; HM-BWT and 3M-BWT.

Fig. 11 compares the circuit latency mapped by Squash 2 and QSPR. As can be seen, Squash 2 could achieve the best results in all of problem sizes for 3M-BWT but it lags behind QSPR for mapping HM-BWT. As it was explained earlier, Squash 2 hides most of the routing delay by parallelizing it with the execution of logical operations. That is why Squash 2 performs better than QSPR when mapping 3M-BWT. On the other hand, HM-BWT suffers from its aggressive modularized structure and the overhead it causes. The overhead includes the architecture reconfiguration delay as well as limited parallelism.

Fig. 12 compares the runtime of QSPR and Squash 2. As can be seen, Squash 2 is always faster than QSPR when mapping HM-BWT. On the other hand, for small problem sizes ($s < 7$), QSPR is slightly faster than Squash 2 in mapping 3M-BWT. However, as the problem size grows, the runtime of QSPR radically increases, whereas the runtime of Squash 2 remains mostly the same. This phenomenon is due to the fact that QSPR handles a large netlist of quantum operations, whereas Squash 2 maps only the quantum modules which grow very slowly compared to the actual circuit size. Moreover, Squash 2 is slower in mapping 3M-BWT compared to HM-BWT because the former has larger modules compared to the latter. Larger modules usually increases the partitioning step runtime. Also note that when $s > 15$, QSPR runtime grows significantly due to the inefficient handling of large netlists.

Table 4 summarizes the latency and runtime of various benchmarks mapped by QSPR and Squash 2. We selected the ancilla budget and core count values that are pointed by the gray arrow in previous subsections for Squash 2. As you may see, in the smallest benchmark, i.e., Grover's Algorithm (n=100), QSPR performs well in terms of the latency. This shows the overhead of modular mapping for small inputs. As the size of Grover's Algorithm increases, the latency of mapped circuits by QSPR gets closer to that of Squash 2. Moreover, QSPR runtime quickly grows, whereas Squash 2 runtime remains mostly the same. As explained before for 3M-BWT, Squash 2 is faster and produces better results compared to QSPR as long as the input is not highly modularized. For the last two algorithms (i.e., *Ground State Estimation* and *Triangle Finding Problem*), QSPR cannot map the input netlist due to their large size. So no fair comparison is possible. This shows the main strength of Squash 2 which enables us to map complex quantum circuits.

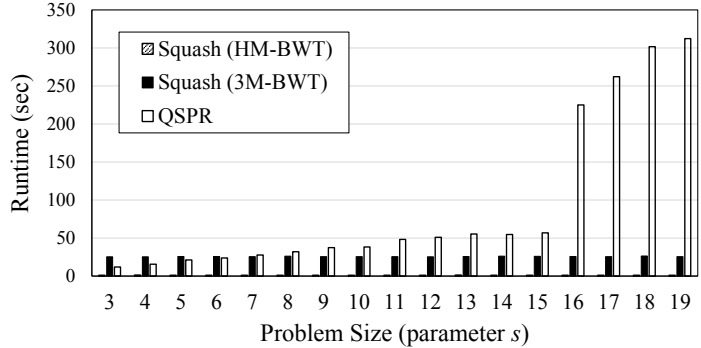

**Fig. 12.** Comparison of QSPR and Squash 2 runtimes for mapping BWT.

**Table 4.** Latency, runtime, and speedup of Squash 2 compared to QSPR for various benchmarks. $m$, $h$, and $y$ are short for minutes, hours, and years, respectively.

| Benchmark | Problem Size | QSPR | | Squash 2 | | Speedup (X) | |
|---|---|---|---|---|---|---|---|
| | | Latency | Runtime (sec) | Latency | Runtime (sec) | Latency | Runtime |
| Grover's Algorithm | n=100 | 3.86m | 8.11 | 4.51m | 0.59 | 0.85 | 13.7 |
| | n=300 | 21.40m | 249 | 24.5m | 0.87 | 0.87 | 285 |
| | n=500 | 47.59m | 1,756 | 53m | 1.09 | 0.90 | 1,608 |
| Binary Welded Tree (3M-BWT) | n=43, s=19 | 187m | 312 | 182m | 25.4 | 1.03 | 12.27 |
| Ground State Estimation | m=6 | N/A[†] | N/A[†] | 44.2h | 1.62 | N/A[†] | N/A[†] |
| Triangle Finding Problem | n=5 | N/A[†] | N/A[†] | 0.92y | 5.99 | N/A[†] | N/A[†] |
| | n=10 | N/A[†] | N/A[†] | 169y | 353 | N/A[†] | N/A[†] |

[†] The QASM file size is very large and QSPR cannot handle it.

## 7 Conclusion

We introduced a hierarchical scalable quantum mapper, called Squash 2. Squash aims to map large quantum algorithms and properly handle the ancilla sharing problem which allows reducing the resource demand. It uses a novel multi-core reconfigurable quantum processor architecture, called Requp, which supports a hierarchical approach to mapping a quantum algorithm and enables ancilla sharing. Experimental results demonstrated that Squash can handle large-scale quantum algorithms while providing an effective mechanism for sharing ancilla qubits.

## Acknowledgements

The authors would like to thank Professor Todd A. Brun for his valuable comments about the calculation of ancilla requirements for various QEC codes and operations.

This research was supported in part by the Intelligence Advanced Research Projects Activity (IARPA) via Department of Interior National Business Center contract number D11PC20165.

## References

[1] M. J. Dousti and M. Pedram, "LEQA: latency estimation for a quantum algorithm mapped to a quantum circuit fabric," in *Proceedings of the Design Automation Conference*, Jun. 2013, pp. 42:1–42:7.

[2] M. G. Whitney, N. Isailovic, Y. Patel, and J. Kubiatowicz, "A fault tolerant, area efficient architecture for Shor's factoring algorithm," in *Proceedings of the International Symposium on Computer Architecture*, Jun. 2009, pp. 383–394.

[3] H. Goudarzi, M. J. Dousti, A. Shafaei, and M. Pedram, "Design of a universal logic block for fault-tolerant realization of any logic operation in trapped-ion quantum circuits," *Quantum Information Processing*, pp. 1267–1299, Jan. 2014.

[4] M. A. Nielsen and I. L. Chuang, *Quantum computation and quantum information*. Cambridge University Press, Dec. 2010.

[5] S. E. Venegas-Andraca, *Quantum walks for computer scientists*. Morgan & Claypool Publishers, 2008.

[6] M. J. Dousti, A. Shafaei, and M. Pedram, "Squash: A scalable quantum mapper considering ancilla sharing," in *Proceedings of the Great Lakes Symposium on VLSI*, May 2014, pp. 117–122.

[7] V. V. Shende and I. L. Markov, "On the CNOT-cost of TOFFOLI gates," *Quantum Information & Computation*, vol. 9, no. 5, pp. 461–486, 2009.

[8] T. S. Metodi, D. D. Thaker, and A. W. Cross, "A quantum logic array microarchitecture: Scalable quantum data movement and computation," in *Proceedings of the International Symposium on Microarchitecture*, Nov. 2005, pp. 305–318.

[9] D. D. Thaker, T. S. Metodi, A. W. Cross, I. L. Chuang, and F. T. Chong, "Quantum memory hierarchies: Efficient designs to match available parallelism in quantum computing," in *Proceedings of the International Symposium on Computer Architecture*, May 2006, pp. 378–390.

[10] N. C. Jones, R. Van Meter, A. G. Fowler, P. L. McMahon, J. Kim, T. D. Ladd, and Y. Yamamoto, "Layered architecture for quantum computing," *Physical Review X*, vol. 2, no. 3, pp. 031 007–1–031 007–27, 2012.

[11] L. Kreger-Stickles and M. Oskin, "Microcoded architectures for ion-trap quantum computers," in *Proceedings of the International Symposium on Computer Architecture*, Jun. 2008, pp. 165–176.

[12] M. J. Dousti and M. Pedram, "Minimizing the latency of quantum circuits during mapping to the ion-trap circuit fabric," in *Proceedings of the Design, Automation, and Test in Europe*, Mar. 2012, pp. 840–843.

[13] N. E. Rainwater and S. J. Kapurch, *NASA Systems Engineering Handbook (NASA/Sp-2007-6105 Rev1)*. National Aeronautics and Space Administration, Dec. 2003.

[14] A. J. Abhari, A. Faruque, M. J. Dousti, L. Svec, O. Catu, A. Chakrabati, C.-F. Chiang, S. Vanderwilt, J. Black, and F. Chong, "Scaffold: Quantum programming language," Department of Computer Science, Princeton University, Tech. Rep. TR-934-12, June 2012.

[15] A. JavadiAbhari, S. Patil, D. Kudrow, J. Heckey, A. Lvov, F. Chong, and M. Martonosi, "ScaffCC: a framework for compilation and analysis of quantum computing programs," in *Proceedings of the Computing Frontiers*, May 2014, pp. 1:1–1:10.

[16] T. D. Ladd, F. Jelezko, R. Laflamme, Y. Nakamura, C. Monroe, and J. L. O'Brien, "Quantum computers," *Nature*, vol. 464, no. 7285, pp. 45–53, 2010.

[17] B. Zeng, A. Cross, and I. Chuang, "Transversality versus universality for additive quantum codes," *IEEE Transactions on Information Theory*, vol. 57, no. 9, pp. 6272–6284, 2011.

[18] M. Tanaka and O. Tatebe, "Workflow scheduling to minimize data movement using multi-constraint graph partitioning," in *Proceedings of the International Symposium on Cluster, Cloud and Grid Computing*, May 2012, pp. 65–72.

[19] G. Karypis and V. Kumar, "Multilevel algorithms for multi-constraint graph partitioning," in *Supercomputing*, Nov. 1998, pp. 1–17.

[20] C.-T. Hwang, J.-H. Lee, and Y.-C. Hsu, "A formal approach to the scheduling problem in high level synthesis," *IEEE Transactions on Computer-Aided Design of Integrated Circuits and Systems*, vol. 10, no. 4, pp. 464–475, 1991.

[21] G. D. Micheli, *Synthesis and optimization of digital circuits*. McGraw-Hill, Jan. 1994.

[22] "Gurobi optimizer," http://www.gurobi.com, [Online; accessed October 5, 2015].

[23] L. K. Grover, "A fast quantum mechanical algorithm for database search," in *Proceedings of the Theory of Computing*, May 1996, pp. 212–219.

[24] A. M. Childs, R. Cleve, E. Deotto, E. Farhi, S. Gutmann, and D. A. Spielman, "Exponential algorithmic speedup by a quantum walk," in *Proceedings of the Theory of Computing*, Jun. 2003, pp. 59–68.

[25] J. D. Whitfield, J. Biamonte, and A. Aspuru-Guzik, "Simulation of electronic structure hamiltonians using quantum computers," *Molecular Physics*, vol. 109, no. 5, pp. 735–750, 2011.

[26] F. Magniez, M. Santha, and M. Szegedy, "Quantum algorithms for the triangle problem," *SIAM Journal on Computing*, vol. 37, no. 2, pp. 413–424, 2007.